\documentclass{article}
\usepackage{graphicx}
\usepackage{amsmath}

\input{tcilatex}

\begin{document}

\title{\textbf{Free Massless Particles, }\\
\textbf{Two Time Physics and}\\
\textbf{Newtonian Gravitodynamics}}
\author{W. Chagas-Filho \\
Departamento de Fisica, Universidade Federal de Sergipe\\
SE, Brazil}
\maketitle

\begin{abstract}
We demonstrate how a classical Snyder-like phase space can be constructed in
the Hamiltonian formalism for the free massless relativistic particle, for
the two-time physics model and for the relativistic Newtonian gravitodynamic
theory. In all these theories the Snyder-like phase space emerges as a
consequence of a new local scale invariance of the Hamiltonian. The
implications and consequences of this Snyder-like phase space in each of
these theories are also considered.
\end{abstract}

\section{Introduction}

Topological and geometrical effects are among the most striking quantum
phenomena discovered since the formulation of quantum mechanics in 1926. The
first of such effects was discovered in 1959 and is termed the Aharonov-Bohm
effect [1]. This effect caused a revolution in the basic assumptions about
the role of potentials in physics. Its discovery prefigured a revolution in
our understanding of gauge fields and of the fundamental forces of nature
[2].

In classical electrodynamics the scalar and vector potentials are no more
than mathematical conveniences, mere auxiliary fields that we can take or
leave. Only the electric and magnetic fields, which act locally on the
charges, are physical fields. In quantum physics the potentials are much
more than mathematical conveniences. It is clear that whenever scalar and
vector potentials appear in the classical Hamiltonian for a physical system,
they will also appear in the corresponding Schrodinger equation, in the
Heisenberg equations and in the Feynman path integral [2]. One of the
results we present in this work is the verification that, in a flat
Euclidean or in a noncommutative space-time, we can have non-vanishing
classical brackets among the canonical variables that describe the motion of
a massless relativistic particle and the components of a relativistic vector
potential $A_{\mu }$ that describes a Newtonian gravitodynamic field. In the
corresponding quantized theory, with these brackets turned into commutators,
there will exist non-vanishing uncertainty relations between the massless
particle's canonical variables and the components of the gravitodynamic
potential. These uncertainty relations are the evidences of the physical
nature of the gravitodynamic potential in quantum mechanics. The brackets we
present in this work are then the evidences of the physical nature of the
gravitodynamic potential in classical physics. Before turning to these
subjects, let us explain the motivation for this work.

There is at the moment a considerable amount of information theorists expect
that a consistent quantum theory of gravity should incorporate. To begin
with, such a theory should incorporated the consequences [3] of the
holographic principle. This principle requires that the density of the
available information about quantum gravity in a certain space-time region
be bounded by the surface surrounding that region. The density of
information on the surface can not exceed one bit of information per Planck
area.

There is now reasonable confidence that quantum gravity should be formulated
in a noncommutative space-time. This comes in as a consequence of additional
quantum uncertainties among the canonical variables that are introduced by
gravitational effects at the Planck length scale [4]. The first attempt to
define a Lorentz invariant noncommutative space-time was performed by Snyder
[5] in 1947. Snyder's noncommutative quantized space-time was originally
proposed as way to solve the ultraviolet divergence problem in quantum field
theory and has recently attracted some interest of researchers working in
quantum gravity and in the physics of mesoscopic systems.

A Snyder-like space-time for quantum gravity can bring new light in the
context of the connection between the infrared divergences of a theory
containing gravity in a certain region of space-time and the ultraviolet
divergences in the conformal field theory living on the boundary of this
region, the so called IR/UV connection, or anti-de Sitter/conformal field
theory (AdS/CFT) connection, which is one of the cornerstones [6] of the
holographic principle. Some time ago, a new de Sitter/Snyder-Yang space-time
connection was proposed [7] in substitution to the AdS/CFT correspondence.
The idea behind this proposal is that conformal field theory entirely lacks
the space-time noncommutativity necessary to make quantum gravity a finite
theory. Snyder-like brackets in $D$ dimensions were recently derived [8] in
the reduced phase space of the $D+2$ dimensional two-time physics model
[9,10,11,12] using the Dirac bracket technique [15] for the quantization of
systems with second class Hamiltonian constraints. This result is
encouraging because the two-time physics model is classically equivalent to
(0+1) dimensional conformal gravity [13,14].

Despite the harmony between all the physical ideas about quantum gravity
described above, a basic difficulty remains unresolved: which physical
object is to be considered as the fundamental bit of information occupying
the Planck area? In this work we support the idea that a relativistic
gravitational dipole at the Planck length can be turned into a possible
candidate for such a fundamental object. As pointed out in [16], the concept
of a free massless relativistic point particle becomes ambiguous in a
noncommutative space-time. Our proposal is that a way out of this difficulty
is to admit that a free massless relativistic particle in a noncommutative
space-time becomes dynamically equivalent to an extended relativistic object
whose spatial extension is a measure of the length scale at which space-time
becomes non-commutative. The simplest possible extended relativistic object
with an effective zero mass is a gravitational dipole, composed of two very
small opposite masses separated by a very small distance. Negative masses
[17] can mathematically appear in massless relativistic particle theory as a
consequence of a non-vanishing minimal space-time length. A brief and rough
discussion of the model of a free massless relativistic particle as a
gravitational dipole at the Planck length can be found in [18].

According to Newtonian gravitation opposite masses should repel each other
and so the dipole structure is unstable, but it can become stable if other
effects are introduced. One of these effects is the nearby presence of many
other identical gravitational dipoles. In this case the many attractive and
repulsive forces between the individual gravitational charges can create
stable dynamical configurations. The initial study [18] suggests that such
stable dynamical configurations of gravitational dipoles can exist in the
tensionless, high energy limit [30], of relativistic string and membrane
theory. The idea that strings should be regarded as composite systems of
more fundamental point-like objects was first introduced by Thorn [31] in
1991. An oscillating gravitational dipole would give rise to time-dependent
gravitoelectric and gravitomagnetic fields and, consequently, to
gravitational waves in empty space.

Gravitomagnetism is a natural prediction of general relativity when
gravitational currents are taken into account. However, the history of
gravitomagnetism begins before the advent of general relativity. Maxwell
himself [19], and some time later Heaviside [20], inspired by the analogy
between Newtonian gravity and electrostatics, dedicated part of their work
in searching an evidence of the existence of gravitomagnetism. Formal recent
developments of gravitomagnetism in the framework of general relativity [21]
and in the non-relativistic Newtonian framework [22] can be found in the
literature. In particular, it was verified in [21] that in the weak field
approximation the gravitomagnetic equation in empty space derived using
general relativity is exactly analogous to Faraday's law of electromagnetic
induction in electrodynamics. In this work we present an initial study of
the classical Hamiltonian formalism for the action describing a massless
relativistic particle moving in a background gravitodynamic potential. Our
contributions to the gravitodynamic theory of this model are: a) we reveal
that this model defines a genuine generally covariant system b) we reveal
that this model defines a conformal theory c) we expose the fact that this
gravitodynamic theory, formulated in a flat Euclidean space-time or in a
noncommutative Snyder-like space-time, have non-vanishing Poisson brackets
between the phase space variables describing the motion of the massless
particle and the components of the vector potential that describes the
background gravitodynamic field, which is an indication of the physical
nature of the potential in the classical theory d) we give a formal
derivation of the non-relativistic equation of motion for a massless
particle moving in a background Newtonian gravitodynamic field. These
subjects will be considered in section four.

As an introduction to the gravitodynamic theory, the paper also contains the
following developments. In section two we present a new scale invariance of
the Hamiltonian that describes a free massless relativistic particle. We
then show how this invariance can be used to perform a transition to a phase
space with Snyder-like brackets. This result acquires importance with
connection to the fact that the massless particle is a prototype of general
relativity and also of string and membrane theories. Section two closes with
an explicit verification that, modulo the Hamiltonian constraint, the Snyder
brackets we derived preserve the usual conformal invariance of the free
massless particle action. One immediate consequence of this verification is
that in the classical free massless particle theory we can use the de
Sitter/Snyder-Yang connection in place of the AdS/CFT connection, as
proposed in [7], with no information loss because conformal invariance in
the Snyder-like classical space-time is restored when the Hamiltonian
constraint is imposed.

As has been demonstrated [11,12], the free massless particle, the harmonic
oscillator and the Hydrogen atom are some of the many dual systems that have
a unified description given by the two-time physics model. For this reason,
it is theoretically important to investigate the existence of the same type
of Snyder-like brackets in the two-time physics formalism. This task is
taken in section three, where it is shown that the same type of scale
invariance we found for the free massless particle Hamiltonian can be found
for the two-time physics Hamiltonian The usual canonical Poisson brackets
and two different types of Snyder-like brackets are then derived using this
invariance. Section three concludes with an explicit verification that one
of the two Snyder-like sets of brackets we derived preserve the Lorentz
invariance of the 2T model. The implications of the other Snyder-like set,
which seems to be related with the relativistic physics of mesoscopic
systems, remains to be investigated.

Section four contains our contributions to the Newtonian gravitodynamic
theory. Gravitomagnetism is turning into a subject of appreciable interest
as a consequence of the rapidly increasing technological refinement in the
measuring and detecting devices used in quantum gravity experiments. It has
been pointed out [21] that gravitomagnetism can explain the anomalous
acceleration observed on the Pioneer spacecraft and also contribute [22] to
the experimentally observed Lense-Thirring effect (dragging of the inertial
frame in a gravitational field). It is then interesting to investigate these
and other effects in the context of a relativistic Newtonian gravitodynamic
theory and section four presents some first steps in this direction.
Concluding remarks appear in section five.

\section{Relativistic Particles}

A relativistic particle describes is space-time a one-parameter trajectory $%
x^{\mu }(\tau )$. A possible form of the action is the one proportional to
the arc length traveled by the particle and given by 
\begin{equation}
S=-m\int ds=-m\int d\tau \sqrt{-\dot{x}^{2}}  \tag{2.1}
\end{equation}
$\tau $ is an arbitrary parameter along the particle's world-line, $m$ is
the particle's mass and $ds^{2}=-\delta _{\mu \nu }dx^{\mu }dx^{\nu }$. We
work in a $D$-dimensional Euclidean space-time with $\mu =1,...,D$. A dot
denotes derivatives with respect to $\tau $ and we use units in which $\hbar
=c=1$.

Action (2.1) is invariant under the Poincar\'{e} transformation 
\begin{equation}
\delta x^{\mu }=a^{\mu }+\omega _{\nu }^{\mu }x^{\nu }  \tag{2.2}
\end{equation}
where $a^{\mu }$ is a constant vector and $\omega _{\mu \nu }=-\omega _{\nu
\mu }$ is a constant matrix. As a consequence of the invariance of action
(2.1) under transformation (2.2), the following field can be defined in
space-time 
\begin{equation}
V=a^{\mu }p_{\mu }-\frac{1}{2}\omega ^{\mu \nu }M_{\mu \nu }  \tag{2.3}
\end{equation}
where $p_{\mu }$ is the particle's momentum and $M_{\mu \nu }=x_{\mu }p_{\nu
}-x_{\nu }p_{\mu }$. Introducing the fundamental Poisson brackets 
\begin{equation}
\{p_{\mu },p_{\nu }\}=0  \tag{2.4a}
\end{equation}
\begin{equation}
\{x_{\mu },p_{\nu }\}=\delta _{\mu \nu }  \tag{2.4b}
\end{equation}
\begin{equation}
\{x_{\mu },x_{\nu }\}=0  \tag{2.4c}
\end{equation}
we find that the generators of the field $V$ obey the algebra 
\begin{equation}
\{p_{\mu },p_{\nu }\}=0  \tag{2.5a}
\end{equation}
\begin{equation}
\{p_{\mu },M_{\nu \lambda }\}=\delta _{\mu \nu }p_{\lambda }-\delta _{\mu
\lambda }p_{\nu }  \tag{2.5b}
\end{equation}
\begin{equation}
\{M_{\mu \nu },M_{\rho \lambda }\}=\delta _{\nu \lambda }M_{\mu \rho
}+\delta _{\mu \rho }M_{\nu \lambda }-\delta _{\nu \rho }M_{\mu \lambda
}-\delta _{\mu \lambda }M_{\nu \rho }  \tag{2.5c}
\end{equation}
This is the Poincar\'{e} space-time algebra in $D$ dimensions. Action (2.1)
is also invariant under the reparametrizations of the world-line 
\begin{equation}
\tau \rightarrow \tau ^{\prime }=f(\tau )  \tag{2.6}
\end{equation}
where $f$ is an arbitrary continuous function of $\tau .$ As a consequence
of its invariance under transformation (2.6), the particle action (2.1)
defines the simplest possible generally covariant physical system.

Action (2.1) is obviously inadequate to study the massless limit of
relativistic particle theory and so we must find an alternative action. Such
an action can be easily computed by treating the relativistic particle as a
constrained system. In the transition to the Hamiltonian formalism, action
(2.1) gives the canonical momentum 
\begin{equation}
p_{\mu }=\frac{m}{\sqrt{-\dot{x}^{2}}}\dot{x}_{\mu }  \tag{2.7}
\end{equation}
and this momentum gives rise to the primary constraint 
\begin{equation}
\phi =\frac{1}{2}(p^{2}+m^{2})=0  \tag{2.8}
\end{equation}
In this work we follow Dirac's [15] convention that a constraint is set
equal to zero only after all calculations have been performed. The canonical
Hamiltonian corresponding to action (2.1), $H=p.\dot{x}-L$, identically
vanishes. Dirac's Hamiltonian for the relativistic particle is then 
\begin{equation}
H_{D}=H+\lambda \phi =\frac{1}{2}\lambda (p^{2}+m^{2})  \tag{2.9}
\end{equation}
where $\lambda (\tau )$ is a Lagrange multiplier, to be interpreted as an
independent variable. We see from (2.8) and (2.9) that the dynamics of the
relativistic particle is not governed by a true Hamiltonian but rather by a
Hamiltonian constraint. The Lagrangian that corresponds to (2.9) is 
\begin{equation}
L=p.\dot{x}-\frac{1}{2}\lambda (p^{2}+m^{2})  \tag{2.10}
\end{equation}
Solving the equation of motion for $\ p_{\mu }$ that \ follows from (2.10)
and inserting the result back in it, we obtain the particle action 
\begin{equation}
S=\int d\tau (\frac{1}{2}\lambda ^{-1}\dot{x}^{2}-\frac{1}{2}\lambda m^{2}) 
\tag{2.11}
\end{equation}
In action (2.11), $\lambda (\tau )$ can be associated [23] to a ``world-line
metric'' $\gamma _{\tau \tau }$ , $\lambda (\tau )=[-\gamma _{\tau \tau
}(\tau )]^{\frac{1}{2}}$ , such that $ds^{2}=\gamma _{\tau \tau }d\tau d\tau 
$. In (2.11), $\lambda (\tau )$ is an ``einbein'' field. In more dimensions,
the ``vielbein'' $e_{\mu }^{a}$ is an alternative description of the metric
tensor. In this context, the particle mass $m$ plays the role of a
(0+1)-dimensional ``cosmological constant''. Action (2.11) is classically
equivalent to action (2.1). This can be checked in the following way. If we
solve the classical equation of motion for $\lambda (\tau )$ that follows
from (2.11) we get the result $\lambda =\pm (\sqrt{-\dot{x}^{2}}/m)$.
Inserting the solution with the positive sign in (2.11), it becomes
identical to (2.1).\ The great advantage of action (2.11) is that it has a
smooth transition to the $m=0$ limit.

The general covariance of action (2.11) manifests itself through invariance
under the transformation 
\begin{equation}
\delta x^{\mu }=\epsilon \dot{x}^{\mu }  \tag{2.12a}
\end{equation}
\begin{equation}
\delta \lambda =\frac{d}{d\tau }(\epsilon \lambda )  \tag{2.12b}
\end{equation}
where $\epsilon (\tau )$ is an arbitrary infinitesimal parameter. Varying $%
x^{\mu }$ in (2.11) we obtain the classical equation for free motion 
\begin{equation}
\frac{d}{d\tau }(\frac{\dot{x}_{\mu }}{\lambda })=\dot{p}_{\mu }=0 
\tag{2.13}
\end{equation}

Now we make a transition to the massless limit. This limit is described by
the action 
\begin{equation}
S=\frac{1}{2}\int d\tau \lambda ^{-1}\dot{x}^{2}  \tag{2.14}
\end{equation}
Action (2.14) is invariant under the Poincar\'{e} transformation (2.2) with $%
\delta \lambda =0$ and under the infinitesimal reparametrization (2.12).
However, action (2.14) exhibits other invariances which are not shared by
the massive particle action (2.11). These are the invariance under global
scale transformations 
\begin{equation}
\delta x^{\mu }=\alpha x^{\mu }  \tag{2.15a}
\end{equation}
\begin{equation}
\delta \lambda =2\alpha \lambda  \tag{2.15b}
\end{equation}
where $\alpha $ is a constant, and invariance under the conformal
transformations 
\begin{equation}
\delta x^{\mu }=(2x^{\mu }x^{\nu }-\delta ^{\mu \nu }x^{2})b_{\nu } 
\tag{2.16a}
\end{equation}
\begin{equation}
\delta \lambda =4\lambda x.b  \tag{2.16b}
\end{equation}
where $b_{\mu }$ is a constant vector. As a consequence of the presence of
these two additional invariances, in the massless particle case the field
(2.3) can be extended to 
\begin{equation}
V=a^{\mu }p_{\mu }-\frac{1}{2}\omega ^{\mu \nu }M_{\mu \nu }+\alpha D+b^{\mu
}K_{\mu }  \tag{2.17}
\end{equation}
with the additional generators 
\begin{equation}
D=x^{\mu }p_{\mu }  \tag{2.18}
\end{equation}
\begin{equation}
K_{\mu }=(2x_{\mu }x^{\nu }-\delta _{\mu }^{\nu }x^{2})p_{\nu }  \tag{2.19}
\end{equation}
which correspond to invariances (2.15) and (2.16), respectively. The
generators of the vector field (2.17) now define the space-time algebra 
\begin{equation}
\{p_{\mu },p_{\nu }\}=0  \tag{2.20a}
\end{equation}
\begin{equation}
\{p_{\mu },M_{\nu \lambda }\}=\delta _{\mu \nu }p_{\lambda }-\delta _{\mu
\lambda }p_{\nu }  \tag{2.20b}
\end{equation}
\begin{equation}
\{M_{\mu \nu },M_{\lambda \rho }\}=\delta _{\nu \lambda }M_{\mu \rho
}+\delta _{\mu \rho }M_{\nu \lambda }-\delta _{\nu \rho }M_{\mu \lambda
}-\delta _{\mu \lambda }M_{\nu \rho }  \tag{2.20c}
\end{equation}
\begin{equation}
\{D,D\}=0  \tag{2.20d}
\end{equation}
\begin{equation}
\{D,p_{\mu }\}=p_{\mu }  \tag{2.20e}
\end{equation}
\begin{equation}
\{D,M_{\mu \nu }\}=0  \tag{2.20f}
\end{equation}
\begin{equation}
\{D,K_{\mu }\}=-K_{\mu }  \tag{2.20g}
\end{equation}
\begin{equation}
\{p_{\mu }.K_{\nu }\}=-2\delta _{\mu \nu }D+2M_{\mu \nu }  \tag{2.20h}
\end{equation}
\begin{equation}
\{M_{\mu \nu },K_{\lambda }\}=\delta _{\nu \lambda }K_{\mu }-\delta
_{\lambda \mu }K_{\nu }  \tag{2.20i}
\end{equation}
\begin{equation}
\{K_{\mu },K_{\nu }\}=0  \tag{2.20j}
\end{equation}
The algebra (2.20) is the conformal space-time algebra [29]. The free
massless particle theory defined by action (2.14) is a conformal theory in $%
D $ space-time dimensions.

The classical equation of motion for $x^{\mu }$ that follows from action
(2.14) is identical to (2.13). The equation of motion for $\lambda $ gives
the condition $\dot{x}^{2}=0$, which tells us that a free massless
relativistic particle moves at the speed of light. As a consequence of this,
it becomes impossible to solve for $\lambda (\tau )$ from its equation of
motion. In the massless theory the value of $\lambda (\tau )$ is completely
arbitrary.

In the transition to the Hamiltonian formalism the massless action (2.14)
gives the canonical momenta 
\begin{equation}
p_{\lambda }=0  \tag{2.21}
\end{equation}
\begin{equation}
p_{\mu }=\frac{\dot{x}_{\mu }}{\lambda }  \tag{2.22}
\end{equation}
and the canonical Hamiltonian 
\begin{equation}
H=\frac{1}{2}\lambda p^{2}  \tag{2.23}
\end{equation}
Equation (2.21) is a primary constraint. Introducing the Lagrange multiplier 
$\xi (\tau )$ for this constraint we can write the Dirac Hamiltonian 
\begin{equation}
H_{D}=\frac{1}{2}\lambda p^{2}+\xi p_{\lambda }  \tag{2.24}
\end{equation}
Requiring the dynamical stability of constraint (2.21), $\dot{p}_{\lambda
}=\{p_{\lambda },H_{D}\}=0$, we obtain the secondary constraint 
\begin{equation}
\phi =\frac{1}{2}p^{2}=0  \tag{2.25}
\end{equation}
Constraint (2.25) has a vanishing Poisson bracket with itself and with
constraint (2.21), being therefore a first-class constraint. As a
consequence, among the $D$ variables $x^{\mu }$ that appear in action (2.14)
only $D-1$ correspond to real physical degrees of freedom. Constraint (2.21)
generates translations in the arbitrary variable $\lambda (\tau )$ and can
therefore be dropped from the formalism.

Now we present a new invariance of the Hamiltonian formalism. The massless
particle Hamiltonian (2.23) is invariant under the transformations 
\begin{equation}
p_{\mu }\rightarrow \tilde{p}_{\mu }=\exp \{-\beta (\dot{x}^{2})\}p_{\mu } 
\tag{2.26a}
\end{equation}
\begin{equation}
\lambda \rightarrow \exp \{2\beta (\dot{x}^{2})\}\lambda  \tag{2.26b}
\end{equation}
where $\beta $ is an arbitrary function of $\dot{x}^{2}$. From the equation
(2.22) for the canonical momentum we find that $x^{\mu }$ should transform
as 
\begin{equation}
x^{\mu }\rightarrow \tilde{x}^{\mu }=\exp \{\beta (\dot{x}^{2})\}x^{\mu } 
\tag{2.27}
\end{equation}
when $p_{\mu }$ transforms as in (2.26a).

Consider now the bracket structure that transformations (2.26a) and (2.27)
induce in the massless particle's phase space. Taking $\beta (\dot{x}%
^{2})=\beta (\lambda ^{2}p^{2})$ in transformations (2.26a) and (2.27), and
retaining only the linear terms in $\beta $ in the exponentials, we find
that the new transformed canonical variables $(\tilde{x}_{\mu },\tilde{p}%
_{\mu })$ obey the brackets 
\begin{equation}
\{\tilde{p}_{\mu },\tilde{p}_{\nu }\}=0  \tag{2.28a}
\end{equation}
\begin{equation}
\{\tilde{x}_{\mu },\tilde{p}_{\nu }\}=(1+\beta )[\delta _{\mu \nu }(1-\beta
)-\{x_{\mu },\beta \}p_{\nu }]  \tag{2.28b}
\end{equation}
\begin{equation}
\{\tilde{x}_{\mu },\tilde{x}_{\nu }\}=(1+\beta )[x_{\mu }\{\beta ,x_{\nu
}\}-x_{\nu }\{\beta ,x_{\mu }\}]  \tag{2.28c}
\end{equation}
written in terms of the old canonical variables. These general brackets obey
the non trivial Jacobi identities $(\tilde{x}_{\mu },\tilde{x}_{\nu },\tilde{%
x}_{\lambda })=0$ \ and \ $(\tilde{x}_{\mu },\tilde{x}_{\nu },\tilde{p}%
_{\lambda })=0$, where 
\begin{equation*}
(a,b,c)=\{\{a,b\},c\}+\{\{b,c\},a\}+\{\{c,a\},b\}
\end{equation*}
As we will see in the following, with a suitable choice for $\beta ,$the
brackets (2.28) lead to extensions of the fundamental Poisson brackets that
can be used to define in a noncommutative Snyder space-time in the quantum
theory..

It turns out that the non-commutative space-time geometry induced by the
brackets (2.28) will be completely determined by the choice $\beta (\dot{x}%
^{2})=\beta (\lambda ^{2}p^{2})=\frac{1}{2}\lambda ^{2}p^{2}$. This is
because classical Poisson brackets, as quantum commutators, satisfy the
property $\{A^{n},B\}=nA^{n-1}\{A,B\}$. If we consider, for instance, $\beta
=(\lambda ^{2}p^{2})^{2}$ and compute $\{\beta ,x^{\mu }\}$ we will find $%
\{\beta ,x^{\mu }\}=2\lambda ^{2}p^{2}\{\lambda ^{2}p^{2},x^{\mu }\}$, and
the right hand side vanishes when constraint (2.25) is imposed. Similarly,
all higher order terms will vanish when (2.25) is imposed, and the
space-time geometry is completely determined by the case $\beta =\frac{1}{2}%
\lambda ^{2}p^{2}.$ Computing the brackets (2.28b) and (2.28c) for this form
of $\beta ,$ and finally imposing constraint (2.25), we arrive at the
brackets 
\begin{equation}
\{\tilde{p}_{\mu },\tilde{p}_{\nu }\}=0  \tag{2.29a}
\end{equation}
\begin{equation}
\{\tilde{x}_{\mu },\tilde{p}_{\nu }\}=\delta _{\mu \nu }-\lambda ^{2}p_{\mu
}p_{\nu }  \tag{2.29b}
\end{equation}
\begin{equation}
\{\tilde{x}_{\mu },\tilde{x}_{\nu }\}=-\lambda ^{2}(x_{\mu }p_{\nu }-x_{\nu
}p_{\mu })  \tag{2.29c}
\end{equation}
while, again imposing constraint (2.25), the transformation equations
(2.26a) and (2.27) become the identity transformation 
\begin{equation}
x^{\mu }\rightarrow \tilde{x}^{\mu }=x^{\mu }  \tag{2.30a}
\end{equation}
\begin{equation}
p_{\mu }\rightarrow \tilde{p}_{\mu }=p_{\mu }  \tag{2.30b}
\end{equation}
Using equations (2.30) in (2.29), we can then write down the brackets 
\begin{equation}
\{p_{\mu },p_{\nu }\}=0  \tag{2.31a}
\end{equation}
\begin{equation}
\{x_{\mu },p_{\nu }\}=\delta _{\mu \nu }-\lambda ^{2}p_{\mu }p_{\nu } 
\tag{2.31b}
\end{equation}
\begin{equation}
\{x_{\mu },x_{\nu }\}=-\lambda ^{2}(x_{\mu }p_{\nu }-x_{\nu }p_{\mu }) 
\tag{2.31c}
\end{equation}
In the transition to the quantum theory the brackets (2.31) will exactly
reproduce the Snyder commutators [5] proposed in 1947 (with $\lambda ^{2}$
playing the role of the noncommutativity parameter $\theta $).

The brackets (2.31) obey all Jacobi identities among the canonical
variables. Since in the massless theory the value of $\lambda (\tau )$ is
arbitrary, as we saw above, we can use the reparametrization invariance
(2.12) to choose a gauge in which $\lambda =1$. We then end up with the
brackets 
\begin{equation}
\{p_{\mu },p_{\nu }\}=0  \tag{2.32a}
\end{equation}
\begin{equation}
\{x_{\mu },p_{\nu }\}=\delta _{\mu \nu }-p_{\mu }p_{\nu }  \tag{2.32b}
\end{equation}
\begin{equation}
\{x_{\mu },x_{\nu }\}=-(x_{\mu }p_{\nu }-x_{\nu }p_{\mu })  \tag{2.32c}
\end{equation}
Computing the conformal algebra using the brackets (2.32) instead of the
Poisson brackets (2.4) we obtain 
\begin{equation}
\{p_{\mu },p_{\nu }\}=0  \tag{2.33a}
\end{equation}
\begin{equation}
\{p_{\mu },M_{\nu \lambda }\}=\delta _{\mu \nu }p_{\lambda }-\delta _{\mu
\lambda }p_{\nu }  \tag{2.33b}
\end{equation}
\begin{equation}
\{M_{\mu \nu },M_{\lambda \rho }\}=\delta _{\nu \lambda }M_{\mu \rho
}+\delta _{\mu \rho }M_{\nu \lambda }-\delta _{\nu \rho }M_{\mu \lambda
}-\delta _{\mu \lambda }M_{\nu \rho }  \tag{2.33c}
\end{equation}
\begin{equation}
\{D,D\}=0  \tag{2.33d}
\end{equation}
\begin{equation}
\{D,p_{\mu }\}=p_{\mu }-p_{\mu }p^{2}  \tag{2.33e}
\end{equation}
\begin{equation}
\{D,M_{\mu \nu }\}=0  \tag{2.33f}
\end{equation}
\begin{equation}
\{D,K_{\mu }\}=-K_{\mu }+2K_{\mu }p^{2}  \tag{2.33g}
\end{equation}
\begin{equation}
\{p_{\mu },K_{\nu }\}=-2\delta _{\mu \nu }D+2M_{\mu \nu }+2p_{\mu }x_{\nu
}p^{2}  \tag{2.33h}
\end{equation}
\begin{equation}
\{M_{\mu \nu },K_{\lambda }\}=\delta _{\nu \lambda }K_{\mu }-\delta
_{\lambda \mu }K_{\nu }  \tag{2.33i}
\end{equation}
\begin{equation}
\{K_{\mu },K_{\nu }\}=2x^{2}M_{\mu \nu }p^{2}  \tag{2.33j}
\end{equation}
Imposing again constraint (2.25) we see that the brackets (2.32) preserve
the structure (2.20) of the $D$ dimensional conformal space-time algebra.
The brackets (2.32) then have a peaceful \ coexistence with conformal
invariance in free massless particle theory. In the next section we will
verify that the brackets (2.32) also preserve the $(D+2)$ dimensional
Lorentz invariance of the two-time physics model.

\section{Two-time Physics}

As we mentioned in the introduction, the free massless particle is one of
the physical systems that have a unified description given by the two-time
physics model. The massless particle Hamiltonian in $D$ dimensions is
obtained by making two gauge choices in the $D+2$ dimensional Hamiltonian
formalism for the 2T model. It is then interesting to investigate the
existence, in 2T physics, of the $D+2$ extension of the $D$ dimensional
brackets (2.32), an extension which is still lacking in the literature. This
is done in this section.

The central idea in two-time physics [9,10,11,12] is to introduce a new
gauge invariance in phase space by gauging the duality of the quantum
commutator $[X_{M},P_{N}]=i\delta _{MN}$. This procedure leads to the
symplectic Sp(2,R) gauge invariance. To remove the distinction between
position and momenta we set $X_{1}^{M}=X^{M}$ and $X_{2}^{M}=P^{M}$ and
define the doublet $X_{i}^{M}=(X_{1}^{M},X_{2}^{M})$. The local Sp(2,R) acts
as 
\begin{equation}
\delta X_{i}^{M}(\tau )=\epsilon _{ik}\omega ^{kl}(\tau )X_{l}^{M}(\tau ) 
\tag{3.1}
\end{equation}
$\omega ^{ij}(\tau )$ is a symmetric matrix containing three local
parameters and $\epsilon _{ij}$ is the Levi-Civita symbol that serves to
raise or lower indices. The Sp(2,R) gauge field $A^{ij}$ is symmetric in $%
(i,j)$ and transforms as 
\begin{equation}
\delta A^{ij}=\partial _{\tau }\omega ^{ij}+\omega ^{ik}\epsilon
_{kl}A^{lj}+\omega ^{jk}\epsilon _{kl}A^{il}  \tag{3.2}
\end{equation}
The covariant derivative is 
\begin{equation}
D_{\tau }X_{i}^{M}=\partial _{\tau }X_{i}^{M}-\epsilon _{ik}A^{kl}X_{l}^{M} 
\tag{3.3}
\end{equation}
An action invariant under the Sp(2,R) gauge symmetry is 
\begin{equation*}
S=\frac{1}{2}\int d\tau (D_{\tau }X_{i}^{M})\epsilon ^{ij}X_{j}^{N}\eta _{MN}
\end{equation*}
After an integration by parts this action can be written as 
\begin{equation*}
=\int d\tau (\partial _{\tau }X_{1}^{M}X_{2}^{N}-\frac{1}{2}%
A^{ij}X_{i}^{M}X_{j}^{N})\eta _{MN}
\end{equation*}
\begin{equation}
=\int d\tau \lbrack \dot{X}.P-(\frac{1}{2}\lambda _{1}P^{2}+\lambda _{2}P.X+%
\frac{1}{2}\lambda _{3}X^{2})]  \tag{3.4}
\end{equation}
where $A^{11}=\lambda _{3}$, $A^{12}=A^{21}=\lambda _{2}$, \ $A^{22}=\lambda
_{1}$ and the canonical Hamiltonian is 
\begin{equation}
H=\frac{1}{2}\lambda _{1}P^{2}+\lambda _{2}P.X+\frac{1}{2}\lambda _{3}X^{2} 
\tag{3.5}
\end{equation}
The equations of motion for the $\lambda $'s give the primary constraints 
\begin{equation}
\phi _{1}=\frac{1}{2}P^{2}=0  \tag{3.6}
\end{equation}
\begin{equation}
\phi _{2}=P.X=0  \tag{3.7}
\end{equation}
\begin{equation}
\phi _{3}=\frac{1}{2}X^{2}=0  \tag{3.8}
\end{equation}
and therefore we can not solve for the $\lambda $'s from their equations of
motion. The values of the $\lambda $'s in action (3.4) are completely
arbitrary. If we consider the Euclidean, or the Minkowski metric, as the
background space, we find that the surface defined by the constraint
equations (3.6)-(3.8) is trivial. The only metric giving a non-trivial
surface and avoiding the ghost problem is the flat metric with two time
coordinates. We then introduce an extra time-like dimension and an extra
space-like dimension and work in a $(D+2)$ dimensional space-time.

We use the fundamental Poisson brackets 
\begin{equation}
\{P_{M},P_{N}\}=0  \tag{3.9a}
\end{equation}
\begin{equation}
\{X_{M},P_{N}\}=\delta _{MN}  \tag{3.9b}
\end{equation}
\begin{equation}
\{X_{M},X_{N}\}=0  \tag{3.9c}
\end{equation}
where $M,N=1,...,D+2$. and verify that constraints (3.6)-(3.8) obey the
algebra 
\begin{equation}
\{\phi _{1},\phi _{2}\}=-2\phi _{1}  \tag{3.10a}
\end{equation}
\begin{equation}
\{\phi _{1},\phi _{3}\}=-\phi _{2}  \tag{3.10b}
\end{equation}
\begin{equation}
\{\phi _{2},\phi _{3}\}=-2\phi _{3}  \tag{3.10c}
\end{equation}
These equations show that all constraints $\phi $ are first class. Equations
(3.10) represent the Lie algebra of the symplectic Sp(2,R) group, which is
the gauge group of the two-time physics model. The gauge algebra (3.10) has
an analogous [8] algebra which is the gauge algebra of the SL(2,R) model
[26]. The SL(2.R) model mimics the gauge symmetry of general relativity and
has the interesting feature that the dynamics is not governed by a true
Hamiltonian but rather by two Hamiltonian constraints. If we redefine the
constraints as $H_{1}=\phi _{1}$ , $H_{2}=-\phi _{3}$ and $D=\phi _{2}$ the
algebra (3.10) becomes 
\begin{equation}
\{H_{1},D\}=-2H_{1}  \tag{3.11a}
\end{equation}
\begin{equation}
\{H_{1},H_{2}\}=D  \tag{3.11b}
\end{equation}
\begin{equation}
\{H_{2},D\}=-2H_{2}  \tag{3.11c}
\end{equation}
which is the gauge algebra of the SL(2,R) model.

Action (3.4) is in first order form. We can go to a second order formalism
by eliminating the canonical momenta. The classical equation of motion for $%
P_{M}$ that follows from action (3.4) has the solution 
\begin{equation}
P_{M}=\frac{1}{\lambda _{1}}(\dot{X}_{M}-\lambda _{2}X_{M})  \tag{3.12}
\end{equation}
Inserting this solution in action (3.4), we obtain the alternative 2T action 
\begin{equation}
S=\int d\tau \lbrack \frac{1}{2\lambda _{1}}(\dot{X}-\lambda _{2}X)^{2}-%
\frac{1}{2}\lambda _{3}X^{2}]  \tag{3.13}
\end{equation}
This form of the action is a generalization of (0+1) dimensional gravity to
(0+1) dimensional conformal gravity [13,14]. An action analogous to (3.13)
was independently derived in [29]. In the canonical formalism for action
(3.13) the constraints (3.6)-(3.8) appear as secondary constraints that are
necessary for the dynamical stability of the primary constraints $p_{\lambda
_{1}}=p_{\lambda _{2}}=p_{\lambda _{3}}=0$. Because constraints (3.6)-(3.8)
are first-class, only $D-1$ of the $D+2$ variables $X_{M}$ that appear in
the 2T action (3.13) correspond to real physical degrees of freedom. The
conformal gravity action (3.13) and the free massless particle action (2.14)
therefore describe the same number of physical degrees of freedom.

Now let us see how the $D+2$ version of the free massless particle brackets
(2.32) can be found in two-time physics. It can be verified that the
Hamiltonian (3.5) is invariant under the transformations 
\begin{equation}
X^{M}\rightarrow \tilde{X}^{M}=\exp \{\beta (\dot{X}^{2})\}X^{M}  \tag{3.14a}
\end{equation}
\begin{equation}
P_{M}\rightarrow \tilde{P}_{M}=\exp \{-\beta (\dot{X}^{2})\}P_{M} 
\tag{3.14b}
\end{equation}
\begin{equation}
\lambda _{1}\rightarrow \exp \{2\beta (\dot{X}^{2})\}\lambda _{1} 
\tag{3.14c}
\end{equation}
\begin{equation}
\lambda _{2}\rightarrow \lambda _{2}  \tag{3.14d}
\end{equation}
\begin{equation}
\lambda _{3}\rightarrow \exp \{-2\beta (\dot{X}^{2})\}\lambda _{3} 
\tag{3.14e}
\end{equation}
$\beta (\dot{X}^{2})$ is an arbitrary function of $\dot{X}^{2}$. The
transformation equations (3.14) are the 2T extensions of the invariance
(2.26), (2.27) we found for the free massless particle. The existence of
this invariance of the 2T Hamiltonian opens the possibility of the existence
of a $D+2$ dimensional extension of the free massless particle brackets
(2.32). But in the 2T model we have an enlarged gauge invariance and
therefore new possibilities.

From the equation (3.12) for the canonical momentum we find 
\begin{equation}
\dot{X}^{2}=\lambda _{1}^{2}P^{2}+2\lambda _{1}\lambda _{2}P.X+\lambda
_{2}^{2}X^{2}  \tag{3.15}
\end{equation}
and in the 2T Hamiltonian formalism the arbitrary function $\beta (\dot{X}%
^{2})$ that appears in transformation (3.14) must be a canonical field $%
\beta (X,P)$. Keeping only the linear terms in $\beta $ in the
transformation (3.14), after some algebra we arrive at the brackets 
\begin{equation}
\{\tilde{P}_{M},\tilde{P}_{N}\}=(\beta -1)[\{P_{M},\beta \}P_{N}+\{\beta
,P_{N}\}P_{M}]+\{\beta ,\beta \}P_{M}P_{N}  \tag{3.16a}
\end{equation}
\begin{equation*}
\{\tilde{X}_{M},\tilde{P}_{N}\}=(1+\beta )[\delta _{MN}(1-\beta
)-\{X_{M},\beta \}P_{N}]
\end{equation*}
\begin{equation}
+(1-\beta )X_{M}\{\beta ,P_{N}\}-X_{M}X_{N}\{\beta ,\beta \}  \tag{3.16b}
\end{equation}
\begin{equation}
\{\tilde{X}_{M},\tilde{X}_{N}\}=(1+\beta )[X_{M}\{\beta
,X_{N}\}-X_{N}\{\beta ,X_{M}\}]+X_{M}X_{N}\{\beta ,\beta \}  \tag{3.16c}
\end{equation}
There are three important choices for $\beta $ and each one leads to a
different set of brackets. Consider first the simplest one, which
corresponds to imposing constraints (3.6)-(3.8) in expression (3.15). We
have $\dot{X}^{2}=0$ and we can then choose $\beta (\dot{X}^{2})=0$ in
expressions (3.16). We then find the brackets 
\begin{equation}
\{P_{M},P_{N}\}=0  \tag{3.17a}
\end{equation}
\begin{equation}
\{X_{M},P_{N}\}=\delta _{MN}  \tag{3.17b}
\end{equation}
\begin{equation}
\{X_{M},X_{N}\}=0  \tag{3.17c}
\end{equation}
where we used the fact that transformations (3.14) become the identity
transformation when $\beta (\dot{X}^{2})=0$. Brackets (3.17) are identical
to the usual Poisson brackets (3.9).

To obtain the extension of the massless particle commutators (2.32) in the
2T model we impose constraints (3.7) and (3.8) in equation (3.15). We can
then choose the form $\beta (\dot{X}^{2})=\frac{1}{2}\lambda _{1}^{2}P^{2}$.
Inserting this form for $\beta $ in the general brackets (3.16), imposing
constraint (3.6) at the end of the calculation so that transformations
(3.14) become the identity transformation, and choosing the gauge $\lambda
_{1}=1$, we arrive at the brackets

\begin{equation}
\{P_{M},P_{N}\}=0  \tag{3.18a}
\end{equation}
\begin{equation}
\{X_{M},P_{N}\}=\delta _{MN}-P_{M}P_{N}  \tag{3.18b}
\end{equation}
\begin{equation}
\{X_{M},X_{N}\}=-(X_{M}P_{N}-X_{N}P_{M})  \tag{3.18c}
\end{equation}
which are the $D+2$ extensions of the free massless particle brackets (2.32).

Another interesting choice is to impose constraints (3.6) and (3.7) first in
expression (3.15). A convenient form for $\beta $ is now $\beta (\dot{X}%
^{2})=\frac{1}{2}\lambda _{2}^{2}X^{2}$. Inserting this form for $\beta $ in
the brackets (3.16), imposing constraint (3.8) at the end of the
calculation, and choosing the gauge $\lambda _{2}=1$, we get the brackets 
\begin{equation}
\{P_{M},P_{N}\}=(X_{M}P_{N}-X_{N}P_{M})  \tag{3.19a}
\end{equation}
\begin{equation}
\{X_{M},P_{N}\}=\delta _{MN}+X_{M}X_{N}  \tag{3.19b}
\end{equation}
\begin{equation}
\{X_{M},X_{N}\}=0  \tag{3.19c}
\end{equation}
which are dual to the brackets (3.18).

The brackets (3.18) can be used to generate a non-linear realization of the
classical Sp(2,R) gauge algebra (3.10). Computing the algebra of the first
class constraints (3.6)-(3.8) using brackets (3.18) we get the new algebra 
\begin{equation}
\{\phi _{1},\phi _{2}\}=-2\phi _{1}+4\phi _{1}^{2}  \tag{3.20a}
\end{equation}
\begin{equation}
\{\phi _{1},\phi _{3}\}=-\phi _{2}+2\phi _{1}\phi _{2}  \tag{3.20b}
\end{equation}
\begin{equation}
\{\phi _{2},\phi _{3}\}=-2\phi _{3}+\phi _{2}^{2}  \tag{3.20c}
\end{equation}
and, by analogy, the gauge algebra (3.11) of the SL(2,R) model becomes 
\begin{equation}
\{H_{1},D\}=-2H_{1}+4H_{1}^{2}  \tag{3.21a}
\end{equation}
\begin{equation}
\{H_{1},H_{2}\}=D-2H_{1}D  \tag{3.21b}
\end{equation}
\begin{equation}
\{H_{2},D\}=-2H_{2}+D^{2}  \tag{3.21c}
\end{equation}
The brackets (3.18) are therefore associated to an enlarged gauge algebra of
the 2T model and of the SL(2,R) model.

At this point it is interesting to consider the $(D+2)$ dimensional Lorentz
invariance of the 2T model. The generators of this symmetry are given by
[11] 
\begin{equation*}
L_{MN}=\epsilon ^{ij}X_{i}^{M}X_{j}^{N}
\end{equation*}
\begin{equation}
=X^{M}P^{N}-X^{N}P^{M}  \tag{3.22}
\end{equation}
and satisfy the bracket 
\begin{equation}
\{L_{MN},L_{RS}\}=\eta _{MR}L_{NS}+\eta _{NS}L_{MR}-\eta _{NR}L_{MS}-\eta
_{MS}L_{NR}  \tag{3.23}
\end{equation}
The generators (3.22) contain the full physical information of the theory
and are gauge invariant [11]. They can therefore be computed in any gauge
with identical results. Following [11] we then choose the basis $%
X_{M}=(X_{+},X_{-},x_{\mu })$ with the metric $\eta _{MN}$ taking the values 
$\eta _{+-}=-1$ and $\eta _{\mu \nu }=\delta _{\mu \nu }$. We choose the two
gauge conditions $X_{+}=1$ and $P_{+}=0$ and solve the two constraints $%
P.X=0 $ and $X^{2}=0$ for $X_{-}$ and $P_{-}$. The final expressions are
[11] 
\begin{equation}
M=(+,-,\mu )  \tag{3.24a}
\end{equation}
\begin{equation}
X_{M}=(1,\frac{1}{2}x^{2},x_{\mu })  \tag{3.24b}
\end{equation}
\begin{equation}
P_{M}=(0,p.x,p_{\mu })  \tag{3.24c}
\end{equation}
The Lorentz generators $L_{MN}$ in this gauge are then given by 
\begin{equation}
L_{++}=L_{--}=0  \tag{3.25a}
\end{equation}
\begin{equation}
L_{+-}=-L_{-+}=p.x  \tag{3.25b}
\end{equation}
\begin{equation}
L_{+\mu }=-L_{\mu +}=p_{\mu }  \tag{3.25c}
\end{equation}
\begin{equation}
L_{-\mu }=-L_{\mu -}=\frac{1}{2}x^{2}p_{\mu }-p.xx_{\mu }  \tag{3.25d}
\end{equation}
\begin{equation}
L_{\mu \nu }=x_{\mu }p_{\nu }-x_{\nu }p_{\mu }  \tag{3.25e}
\end{equation}
Using the $D$ dimensional brackets (2.32) to compute the $D+2$ dimensional
Lorentz algebra generated by these $L_{MN}$ we find 
\begin{equation}
\{L_{+\mu },L_{+\nu }\}=0  \tag{3.26a}
\end{equation}
\begin{equation}
\{L_{+\nu },L_{\nu \lambda }\}=\delta _{\mu \nu }L_{+\lambda }-\delta _{\mu
\lambda }L_{+\nu }  \tag{3.26b}
\end{equation}
\begin{equation}
\{L_{\mu \nu },L_{\nu \rho }\}=\delta _{\nu \lambda }L_{\mu \rho }+\delta
_{\mu \rho }L_{\nu \lambda }-\delta _{\nu \rho }L_{\mu \lambda }-\delta
_{\mu \lambda }L_{\nu \rho }  \tag{3.26c}
\end{equation}
\begin{equation}
\{L_{+-},L_{+-}\}=0  \tag{3.26d}
\end{equation}
\begin{equation}
\{L_{+-},L_{+\mu }\}=L_{+\mu }-L_{+\mu }p^{2}  \tag{3.26e}
\end{equation}
\begin{equation}
\{L_{+-},L_{\mu \nu }\}=0  \tag{3.26f}
\end{equation}
\begin{equation}
\{L_{+-},L_{-\mu }\}=-L_{-\mu }+2L_{-\mu }p^{2}  \tag{3.26g}
\end{equation}
\begin{equation}
\{L_{+\mu },L_{-\nu }\}=\delta _{\mu \nu }L_{+-}-L_{\mu \nu }-p_{\mu }x_{\nu
}p^{2}  \tag{3.26h}
\end{equation}
\begin{equation}
\{L_{\mu \nu },L_{-\lambda }\}=\delta _{\nu \lambda }L_{-\mu }-\delta
_{\lambda \mu }L_{-\nu }  \tag{3.26i}
\end{equation}
\begin{equation}
\{L_{-\mu },L_{-\nu }\}=\frac{1}{2}x^{2}L_{\mu \nu }p^{2}  \tag{3.26j}
\end{equation}
where constraint $\frac{1}{2}p^{2}=0$ must still be imposed. The $D+2$
dimensional Lorentz algebra (3.26) of the two-time physics model becomes
identical to the $D$ dimensional conformal algebra (2.33) of the free
massless particle if we make the identifications 
\begin{equation}
L_{+\mu }=p_{\mu }  \tag{3.27a}
\end{equation}
\begin{equation}
L_{\mu \nu }=M_{\mu \nu }  \tag{3.27b}
\end{equation}
\begin{equation}
L_{+-}=D  \tag{3.27c}
\end{equation}
\begin{equation}
L_{-\mu }=-\frac{1}{2}K_{\mu }  \tag{2.27d}
\end{equation}
We then conclude that, again modulo the constraint, the Snyder brackets
(2.32) also preserve the $D+2$ dimensional Lorentz algebra of the two-time
physics model, and that this algebra is identical to the $D$ dimensional
conformal algebra (2.33) of the free massless relativistic particle. This
result is well known [12,29] to be true on the basis of the fundamental
Poisson brackets (2.4). Now we have proved that it is also true on the basis
of the Snyder brackets (2.32).

\section{Newtonian Gravitodynamics}

The Maxwell-Heaviside equations for Newtonian gravitodynamics were recently
considered in [22]. They are given by 
\begin{equation}
\vec{\nabla}\cdot \mathbf{E}_{g}=-4\pi G\rho  \tag{4.1a}
\end{equation}
\begin{equation}
\vec{\nabla}\times \mathbf{E}_{g}=-\frac{\partial \mathbf{B}_{g}}{\partial t}
\tag{4.1b}
\end{equation}
\begin{equation}
\vec{\nabla}\cdot \mathbf{B}_{g}=0  \tag{4.1c}
\end{equation}
\begin{equation}
\vec{\nabla}\times \mathbf{B}_{g}=-\frac{4\pi G}{c_{g}^{2}}\mathbf{j}+\frac{1%
}{c_{g}^{2}}\frac{\partial \mathbf{E}_{g}}{\partial t}  \tag{4.1d}
\end{equation}
In these equations $G$ is Newton's constant, $\rho $ is a mass density, $%
\mathbf{j}$ is the vector density of mass currents, $\mathbf{E}_{g}$ is the
gravitoelectric vector field given by 
\begin{equation}
\mathbf{E}_{g}=-\vec{\nabla}\Phi -\frac{\partial \mathbf{A}}{\partial t} 
\tag{4.2}
\end{equation}
and $\vec{B}_{g}$ is the gravitomagnetic vector field 
\begin{equation}
\mathbf{B}_{g}=\vec{\nabla}\times \mathbf{A}  \tag{4.3}
\end{equation}
$\Phi $ is the gravitational scalar potential and $\mathbf{A}$ is the
gravitational vector potential for this theory. For details and a brief
review of the history of gravitodynamics see [22] and cited references.

Equations (4.1) contain the unknown constant $c_{g}$ , which is the velocity
of propagation of gravitational waves in empty space. Reasoning by analogy
with electrodynamics, the author in [22] suggests that the force equation
for a particle of mass $m$ moving with velocity $\mathbf{v}$ in a
gravitodynamic field should be 
\begin{equation}
\mathbf{F}=m\mathbf{E}_{g}+m\mathbf{v}\times \mathbf{B}_{g}  \tag{4.4}
\end{equation}
In this section we show how a formalism analogous to the one developed in
the previous two sections can be constructed in the case of a massless
relativistic particle moving in a background gravitodynamic field. We also
derive the massless limit of the non-relativistic equation (4.4).

In the same way as Maxwell's electrodynamics [24], the gravitodynamic field
can be described with the help of a relativistic vector potential $A_{\mu }=(%
\mathbf{A,}i\Phi )$. We can then introduce the $D$ dimensional action 
\begin{equation}
S=\int d\tau (\frac{1}{2}\lambda ^{-1}\dot{x}^{2}+A.\dot{x})  \tag{4.5}
\end{equation}
for a massless relativistic particle moving in an external gravitodynamic
field $A_{\mu }=A_{\mu }(\tau )$. Action (4.5) is invariant under the
infinitesimal reparametrization 
\begin{equation}
\delta x^{\mu }=\epsilon \dot{x}^{\mu }  \tag{4.6a}
\end{equation}
\begin{equation}
\delta \lambda =\frac{d}{d\tau }(\epsilon \lambda )  \tag{4.6b}
\end{equation}
\begin{equation}
\delta A_{\mu }=\epsilon \dot{A}_{\mu }  \tag{4.6c}
\end{equation}
where $\epsilon (\tau )$ is an arbitrary parameter, because the Lagrangian
in action (4.5) transforms as a total derivative, $\delta L=\frac{d}{d\tau }%
(\epsilon L)$, under transformations (4.6). Action (4.5) therefore describes
a genuine generally covariant physical system.

Action (4.5) is also invariant under the Poincar\'{e} transformation 
\begin{equation}
\delta x^{\mu }=a^{\mu }+\omega _{\nu }^{\mu }x^{\nu }  \tag{4.7a}
\end{equation}
\begin{equation}
\delta A^{\mu }=b^{\mu }+\omega _{\nu }^{\mu }A^{\nu }  \tag{4.7b}
\end{equation}
\begin{equation}
\delta \lambda =0  \tag{4.7c}
\end{equation}
under the scale transformation 
\begin{equation}
\delta x^{\mu }=\alpha x^{\mu }  \tag{4.8a}
\end{equation}
\begin{equation}
\delta A^{\mu }=-\alpha A^{\mu }  \tag{4.8b}
\end{equation}
\begin{equation}
\delta \lambda =2\alpha \lambda  \tag{4.8c}
\end{equation}
and under the conformal transformation 
\begin{equation}
\delta x^{\mu }=(2x^{\mu }x^{\nu }-\delta ^{\mu \nu }x^{2})b_{\nu } 
\tag{4.9a}
\end{equation}
\begin{equation}
\delta A^{\mu }=2(x^{\mu }A^{\nu }-A^{\mu }x^{\nu }-A.x\delta ^{\mu \nu
})b_{\nu }  \tag{4.9b}
\end{equation}
\begin{equation}
\delta \lambda =4\lambda x.b  \tag{4.9c}
\end{equation}
The generally covariant gravitodynamic action (4.5) therefore defines a $D$
dimensional conformal theory.

The physically relevant canonical momenta that follow from action (4.5) are 
\begin{equation}
p_{\lambda }=0  \tag{4.10}
\end{equation}
\begin{equation}
p_{\mu }=\frac{\dot{x}_{\mu }}{\lambda }+A_{\mu }  \tag{4.11}
\end{equation}
and the corresponding Hamiltonian is 
\begin{equation}
H=\frac{1}{2}\lambda (p_{\mu }-A_{\mu })^{2}  \tag{4.12}
\end{equation}
Equation (4.10) is a primary constraint. Introducing the Lagrange multiplier 
$\chi (\tau )$ for this constraint, constructing the Dirac Hamiltonian $%
H_{D}=H+\chi p_{\lambda }$ and requiring the dynamical stability of (4.10),
we obtain the secondary constraint 
\begin{equation}
\phi =\frac{1}{2}(p_{\mu }-A_{\mu })^{2}=0  \tag{4.13}
\end{equation}

Now, the gravitodynamic Hamiltonian (4.12) is invariant under the
transformations 
\begin{equation}
p_{\mu }\rightarrow \tilde{p}_{\mu }=\exp \{-\beta (\dot{x}^{2})\}p_{\mu } 
\tag{4.14a}
\end{equation}
\begin{equation}
\lambda \rightarrow \exp \{2\beta (\dot{x}^{2})\}\lambda  \tag{4.14b}
\end{equation}
\begin{equation}
A_{\mu }\rightarrow \tilde{A}_{\mu }=\exp \{-\beta (\dot{x}^{2})\}A_{\mu } 
\tag{4.14c}
\end{equation}
and from expression (4.11) we find that $x^{\mu }$ should transform as 
\begin{equation}
x^{\mu }\rightarrow \tilde{x}^{\mu }=\exp \{\beta (\dot{x}^{2})\}x^{\mu } 
\tag{4.14d}
\end{equation}
when $p_{\mu }$ transforms as in (4.14a). From (4.11) we also have 
\begin{equation}
\dot{x}^{2}=\lambda ^{2}(p^{2}-2p.A+A^{2})  \tag{4.15}
\end{equation}
To proceed in a consistent way it is necessary to introduce, in addition to
the brackets (2.4), the new brackets 
\begin{equation}
\{A_{\mu },x_{\nu }\}=A_{\mu }x_{\nu }  \tag{4.16a}
\end{equation}
\begin{equation}
\{A_{\mu },p_{\nu }\}=-p_{\nu }A_{\mu }  \tag{4.16b}
\end{equation}
\begin{equation}
\{A_{\mu },A_{\nu }\}=0  \tag{4.16c}
\end{equation}
The introduction of bracket (4.16b) will be justified later when we derive
the massless limit of (4.4). The introduction of bracket (4.16a) is
justified by the consistency of the formalism developed below. Brackets
(4.16a) and (4.16b) are indications of the physical nature of the
gravitodynamic potential $A_{\mu }$ in classical physics.

Transformations (4.14) induce the general bracket structure in the
transformed phase space 
\begin{equation}
\{\tilde{p}_{\mu },\tilde{p}_{\nu }\}=(\beta -1)[\{p_{\mu },\beta \}p_{\nu
}+\{\beta ,p_{\nu }\}p_{\mu }]+\{\beta ,\beta \}p_{\mu }p_{\nu }  \tag{4.17a}
\end{equation}
\begin{equation*}
\{\tilde{x}_{\mu },\tilde{p}_{\nu }\}=(1+\beta )[\delta _{\mu \nu }(1-\beta
)-\{x_{\mu },\beta \}p_{\nu }]
\end{equation*}
\begin{equation}
+(1-\beta )x_{\mu }\{\beta ,p_{\nu }\}-x_{\mu }x_{\nu }\{\beta ,\beta \} 
\tag{4.17b}
\end{equation}
\begin{equation}
\{\tilde{x}_{\mu },\tilde{x}_{\nu }\}=(1+\beta )[x_{\mu }\{\beta ,x_{\nu
}\}-x_{\nu }\{\beta ,x_{\mu }\}]+\{\beta ,\beta \}x_{\mu }x_{\nu } 
\tag{4.17c}
\end{equation}
\begin{equation}
\{\tilde{A}_{\mu },\tilde{x}_{\nu }\}=(1-\beta )\{A_{\mu },\beta \}x_{\nu
}-(1+\beta )\{\beta ,x_{\nu }\}A_{\mu }-\{\beta ,\beta \}A_{\mu }x_{\nu } 
\tag{4.17d}
\end{equation}
\begin{equation}
\{\tilde{A}_{\mu },\tilde{p}_{\nu }\}=-(1+2\beta +2\beta ^{2})p_{\nu }A_{\mu
}-(1-2\beta )\{A_{\mu },\beta \}p_{\nu }-A_{\mu }\{\beta ,p_{\nu }\} 
\tag{4.17e}
\end{equation}
\begin{equation}
\{\tilde{A}_{\mu },\tilde{A}_{\nu }\}=0  \tag{4.17f}
\end{equation}
From (4.15), and using the reparametrization invariance (4.6) to impose the
gauge $\lambda ^{2}=\frac{1}{2}$, we can choose the arbitrary function $%
\beta (\dot{x}^{2})$ that appears in transformations (4.14) to be 
\begin{equation}
\beta (\dot{x}^{2})=\frac{1}{2}p^{2}-p.A+\frac{1}{2}A^{2}  \tag{4.18}
\end{equation}
Computing the brackets (4.17) for this form of $\beta $ in terms of the
fundamental brackets (2.4) and (4.16), and imposing constraint (4.13) at the
end of the calculation, we arrive at the brackets 
\begin{equation}
\{\tilde{p}_{\mu },\tilde{p}_{\nu }\}=0  \tag{4.19a}
\end{equation}
\begin{equation}
\{\tilde{x}_{\mu },\tilde{p}_{\nu }\}=\delta _{\mu \nu }-(p_{\mu }-A_{\mu
})p_{\nu }  \tag{4.19b}
\end{equation}
\begin{equation}
\{\tilde{x}_{\mu },\tilde{x}_{\nu }\}=-[x_{\mu }(p_{\nu }-A_{\nu })-x_{\nu
}(p_{\mu }-A_{\mu })]  \tag{4.19c}
\end{equation}
\begin{equation}
\{\tilde{A}_{\mu },\tilde{x}_{\nu }\}=(p_{\nu }-A_{\nu })A_{\mu } 
\tag{4.19d}
\end{equation}
\begin{equation}
\{\tilde{A}_{\mu },\tilde{p}_{\nu }\}=-p_{\nu }A_{\mu }  \tag{4.19e}
\end{equation}
\begin{equation}
\{\tilde{A}_{\mu },\tilde{A}_{\nu }\}=0  \tag{4.19f}
\end{equation}
Imposing now constraint (4.13) in the transformations (4.14a), (4.14c) and
(4.14d), these become the identity transformations 
\begin{equation}
p_{\mu }\rightarrow \tilde{p}_{\mu }=p_{\mu }  \tag{4.20a}
\end{equation}
\begin{equation}
A_{\mu }\rightarrow \tilde{A}_{\mu }=A_{\mu }  \tag{4.20b}
\end{equation}
\begin{equation}
x_{\mu }\rightarrow \tilde{x}_{\mu }=x_{\mu }  \tag{4.20c}
\end{equation}
We then end with the brackets 
\begin{equation}
\{p_{\mu },p_{\nu }\}=0  \tag{4.21a}
\end{equation}
\begin{equation}
\{x_{\mu },p_{\nu }\}=\delta _{\mu \nu }-(p_{\mu }-A_{\mu })p_{\nu } 
\tag{4.21b}
\end{equation}
\begin{equation}
\{x_{\mu },x_{\nu }\}=-[x_{\mu }(p_{\nu }-A_{\nu })-x_{\nu }(p_{\mu }-A_{\mu
})]  \tag{4.21c}
\end{equation}
\begin{equation}
\{A_{\mu },x_{\nu }\}=(p_{\nu }-A_{\nu })A_{\mu }  \tag{4.21d}
\end{equation}
\begin{equation}
\{A_{\mu },p_{\nu }\}=-p_{\nu }A_{\mu }  \tag{4.21e}
\end{equation}
\begin{equation}
\{A_{\mu },A_{\nu }\}=0  \tag{4.21f}
\end{equation}
which extend the free massless particle brackets (2.32) to the case where
the massless particle interacts with a background gravitodynamic potential $%
A_{\mu }(\tau )$. Notice that bracket (4.21c), which will determine the
space-time geometry in the quantum theory, is consistent with the minimal
coupling prescription $p_{\mu }\rightarrow p_{\mu }-A_{\mu }$ applied to the
free bracket (2.32c). The brackets (4.21d) and (4.21e) are again indications
of the physical nature of the gravitodynamic potential in classical physics,
but now in a Snyder-like space-time.

Let us now derive the massless limit of \ the non-relativistic equation
(4.4). To do this we introduce an explicit dependence of the potential $%
A_{\mu }(\tau )$ on the particle's coordinates and rewrite it as $A_{\mu
}(x^{\nu }(\tau )).$ In this case $\{A_{\mu },p_{\nu }\}=\frac{\partial
A_{\mu }}{\partial x^{\nu }}=-p_{\nu }A_{\mu }$ and this explains the
bracket (4.16b) above. With this explicit dependence we obtain from action
(4.5) the equation of motion 
\begin{equation}
\dot{p}_{\mu }=\frac{\partial A_{\nu }}{\partial x^{\mu }}\dot{x}^{\nu }-%
\dot{A}_{\mu }  \tag{4.19}
\end{equation}
The spatial part of this equation reads 
\begin{equation}
\mathbf{\dot{p}=}\vec{\nabla}(\mathbf{A\cdot \dot{r}}-\Phi \dot{t})-\mathbf{%
\dot{A}}  \tag{4.20}
\end{equation}
We can now use the reparametrization invariance (4.6) to choose the gauge $%
\tau =t$. Equation (4.20) then becomes 
\begin{equation}
\frac{d\mathbf{p}}{dt}=\vec{\nabla}(\mathbf{A\cdot v)-}\vec{\nabla}\Phi -%
\frac{d\mathbf{A}}{dt}  \tag{4.21}
\end{equation}
Using the formula $\vec{\nabla}(\mathbf{a\cdot b)=(a\cdot }\vec{\nabla})%
\mathbf{b+(b\cdot }\vec{\nabla})\mathbf{a}+\mathbf{a}\times (\vec{\nabla}%
\times \mathbf{b})\mathbf{+b\times }(\vec{\nabla}\times \mathbf{a})$ and
taking derivatives with respect to the coordinates keeping the velocities
fixed, we find that 
\begin{equation}
\vec{\nabla}(\mathbf{A\cdot v)=(v\cdot }\vec{\nabla})\mathbf{A+v\times (}%
\vec{\nabla}\mathbf{\times A)}  \tag{4.22}
\end{equation}
Substituting this relation in (4.21), it becomes 
\begin{equation}
\frac{d\mathbf{p}}{dt}=(\mathbf{v\cdot }\vec{\nabla})\mathbf{A+v\times (}%
\vec{\nabla}\times \mathbf{A)-}\vec{\nabla}\Phi -\frac{d\mathbf{A}}{dt} 
\tag{4.23}
\end{equation}
But from vector analysis we now that 
\begin{equation}
\frac{d\mathbf{A}}{dt}=\frac{\partial \mathbf{A}}{\partial t}+\mathbf{%
(v\cdot }\vec{\nabla})\mathbf{A}  \tag{4.24}
\end{equation}
and therefore 
\begin{equation}
-\frac{d\mathbf{A}}{dt}+(\mathbf{v\cdot }\vec{\nabla})\mathbf{A=-}\frac{%
\partial \mathbf{A}}{\partial t}  \tag{4.25}
\end{equation}
Now substituting this in (4.23), we obtain 
\begin{equation}
\frac{d\mathbf{p}}{dt}=-\vec{\nabla}\Phi -\frac{\partial \mathbf{A}}{%
\partial t}+\mathbf{v\times (}\vec{\nabla}\times \mathbf{A)}  \tag{4.26}
\end{equation}
With the identifications (4.2) and (4.3), equation (4.26) can be rewritten
as 
\begin{equation}
\frac{d\mathbf{p}}{dt}=\mathbf{E}_{g}+\mathbf{v\times B}_{g}  \tag{4.27}
\end{equation}
which extends the validity of equation (4.4) to the case of massless
particles.

\section{Concluding remarks}

As is well known, in the canonical quantization procedure for a classical
theory, one of the steps to the quantum theory is to define the fundamental
commutators among the quantum operators that will describe the dynamics. The
recipe is that these commutators are defined by multiplying by an ``i'' the
values of the corresponding classical Poisson brackets. The results of this
work then present an alternative possibility of constructing the quantum
mechanics of the free massless relativistic particle, of the two-time
physics model and of the Newtonian gravitodynamic theory. In these
alternative formulations of the quantized theory the fundamental commutator
relations are not of the Heisenberg type but are Snyder commutators.

In each of the above mentioned theories there are important reasons for
investigating a quantum mechanics based on Snyder commutators. The free
massless particle is only one of the many dual physical systems that have a
unified description given by the two-time physics model. The list includes
the harmonic oscillator, the Hydrogen atom, the particle moving in a de
Sitter space, the particle moving in arbitrary attractive and repulsive
potentials and possibly others still unknown. As for the free massless
particle, Snyder brackets may also be hidden in the dynamics of all these
dual physical systems, possibly in different space-time dimensions. As we
saw in section three, the Snyder brackets for the two-time physics model
enlarge the duality gauge algebra of the model and consequently may also
enlarge the number of dual systems it can describe, while still preserving
its $D+2$ dimensional Lorentz invariance. As we saw in this work, the
gravitodynamic theory brings with it a clear comprehension of the
gravitational Aharonov-Bohm effect (also called the Aharonov-Carmi effect
[28]). However, Snyder brackets for the gravitodynamic theory open the
possibility of the existence of entirely new and unexpected gravitational
effects, with no parallel in electrodynamics. Above all advances,
relativistic quantum gravitodynamics in a noncommutative space-time can
bring with it a clue of which is the fundamental physical object occupying
the Planck area.

\end{document}